\documentclass[a4paper]{jpconf}
\usepackage[utf8]{inputenc} 
\usepackage{amsmath}
\usepackage{color} 
\usepackage{hyperref} 
\usepackage{xspace}
\usepackage{amsfonts} 
\usepackage{amssymb} 
\usepackage{graphicx}

\newcommand{\dd}{{\rm d}} 

\newcommand{\UNIT}[1]{\ensuremath{\,{\rm #1}}\xspace}

\newcommand{\MeV}{\UNIT{MeV}} \newcommand{\GeV}{\UNIT{GeV}}
 
\newcommand{\GeVfmc}{\UNIT{GeV/fm^3}} \newcommand{\TeV}{\UNIT{TeV}}
 
 \newcommand{\fm}{\UNIT{fm}}
 
\newcommand{\fmc}{\ensuremath{\,{\rm fm}/c}\xspace}

\definecolor{magenta}{cmyk}{0,1,0,0}

\newcommand{\REM}[1]{}

\begin{document}

\title{Equilibration of hadrons in HICs via Hagedorn States}

\author{M Beitel, K Gallmeister and C Greiner}
\address{Institut f\"ur Theoretische Physik, Goethe-Universit\"at
  Frankfurt am Main, Max-von-Laue-Str.~1, 60438 Frankfurt am Main,
  Germany}
\ead{gallmeister@th.physik.uni-frankfurt.de}

\begin{abstract}
Hagedorn states (HS) are a tool to model the hadronization process which occurs in the phase transition region between the quark gluon plasma (QGP) and the hadron resonance gas (HRG). These states are believed to appear near the Hagedorn temperature $T_H$ which in our understanding equals the critical temperature $T_c$. A covariantly formulated bootstrap equation is solved to generate the zoo of these particles characterized baryon number $B$, strangeness $S$ and electric charge $Q$.
These hadron-like resonances are characterized by being very massive and by not being limited to quantum numbers of known hadrons.  All hadronic properties like masses, spectral functions etc.~are taken from the hadronic transport model Ultra Relativistic Quantum Molecular Dynamics (UrQMD).
Decay chains of single Hagedorn states provide a well description of experimentally observed multiplicity ratios of strange and multi-strange particles. In addition, the final energy spectra of resulting hadrons show a thermal-like distribution with the characteristic Hagedorn temperature $T_H$.
Box calculations including these Hagedorn states are
performed. Indeed, the time scales leading to equilibration of the
system are drastically reduced down to 2\dots5\fmc.
\end{abstract}

\section{Introduction}

Back in the 60's of the last century and before the advent of quantum chromodynamics (QCD) as the theory of strong interactions, R.~Hagedorn 
\cite{Hagedorn:1965st} proposed the existence of a whole zoo of massive, unobserved hadronic resonance states. The spectrum of these particles, known as Hagedorn spectrum, exhibits the specific feature of being exponential in the infinite mass limit with the slope given by the so called Hagedorn temperature $T_H$. This temperature denotes the limiting temperature for hadronic matter since any partition function of a HRG with Hagedorn-like mass spectrum diverges as long as $T>T_H$. Above the Hagedorn temperature a new state of matter, namely the QGP, shall be realized. 
Thus, Hagedorn states provide a tool to understand the phase transition from HRG to QGP and back.
The Hagedorn states are created in multi-particle collisions most abundantly near $T_H$ which in our understanding equals to the critical 
temperature $T_c$. Hagedorn states are color neutral objects which are allowed to have any quantum numbers as long as they are compatible with its mass. The appearance of Hagedorn states in multi-particle collisions and their role was already discussed in 
\cite{Greiner:2000tu,Greiner:2004vm,NoronhaHostler:2007jf,NoronhaHostler:2010dc,
NoronhaHostler:2009cf}.
Their appearance near $T_c$ can explain,
as shown in \cite{NoronhaHostler:2007jf,NoronhaHostler:2010dc,NoronhaHostler:2009cf},
the fast chemical equilibration of (multi-) strange baryons $B$ and their anti-particles $\bar{B}$ at 
Relativistic Heavy Ion Collider (RHIC) energies.
The inclusion of Hagedorn states in a hadron resonance gas model provides
also a lowering of the speed of sound, $c_s$ and of the shear viscosity over entropy density ratio 
$\eta/s$ at the phase transition and
being in good agreement with lattice calculations
\cite{NoronhaHostler:2008ju,Majumder:2010ik,NoronhaHostler:2012ug,Jakovac:2013iua}.

\begin{figure}
  \centering

  \hspace*{\fill}
  \includegraphics[width=5cm,angle=270]{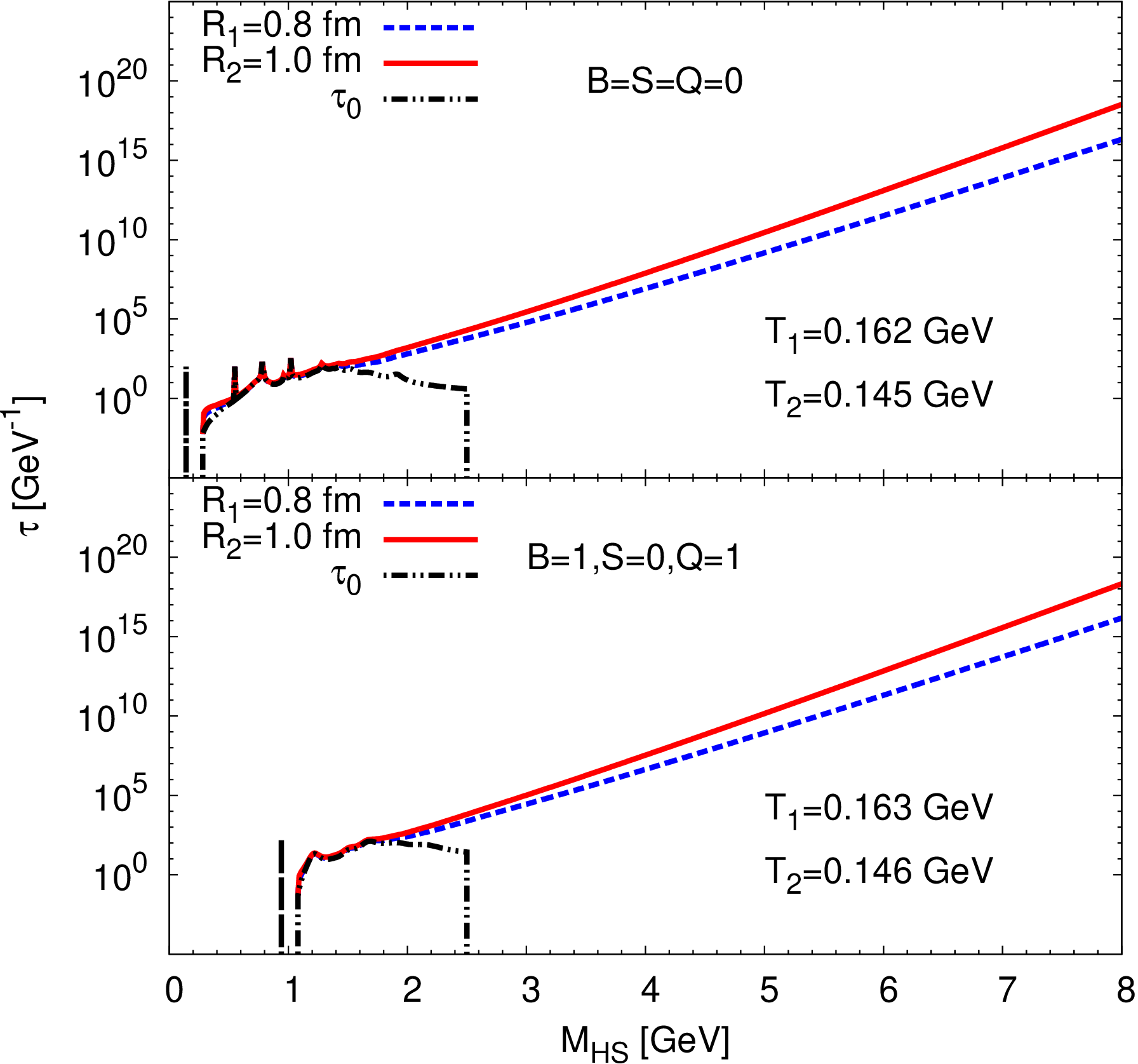}
  \hspace*{\fill}
  \includegraphics[width=5cm,angle=270]{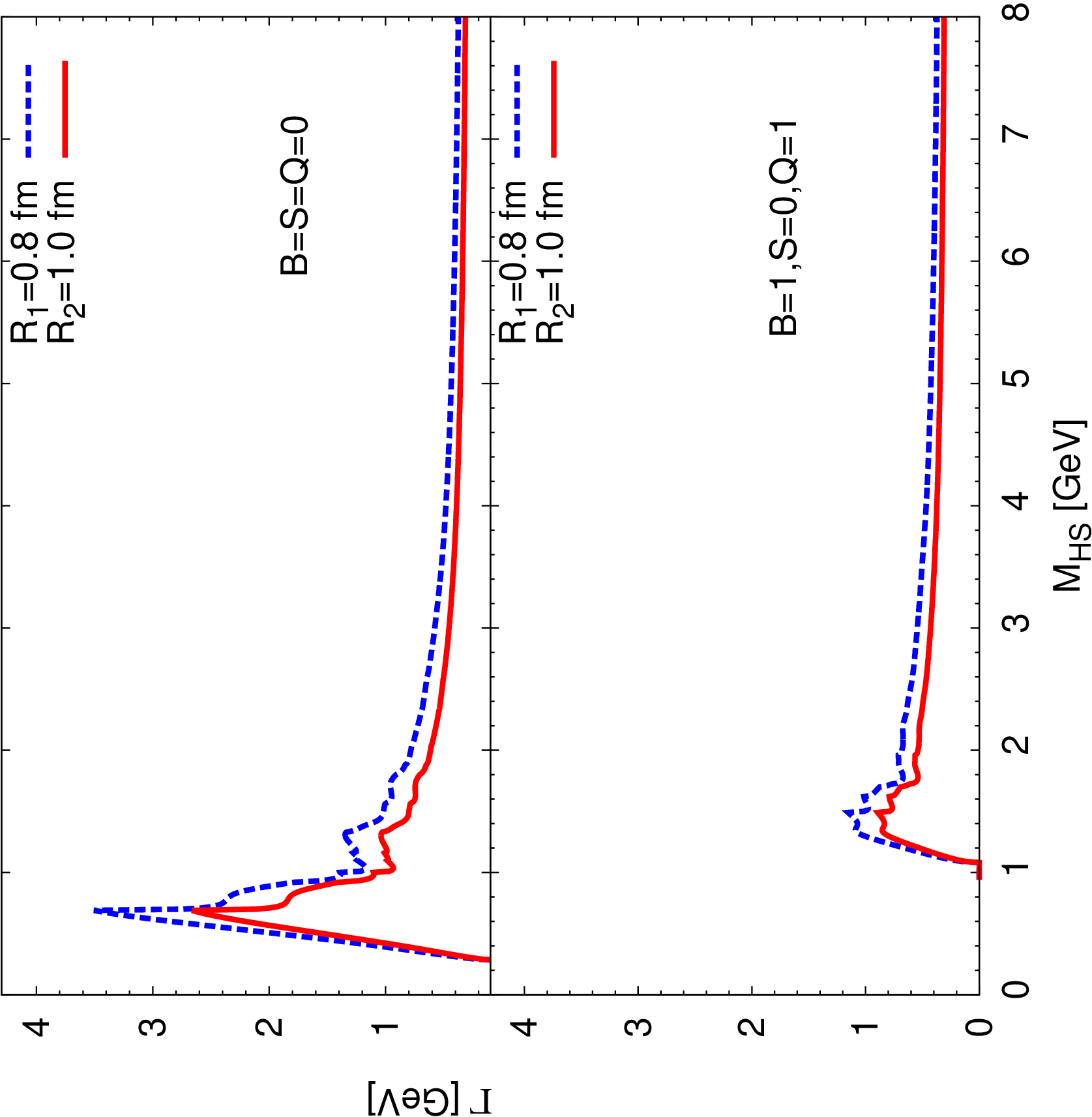}
  \hspace*{\fill}

  \caption{Hagedorn spectrum (left) and total decay width (right) for meson-like $\left(B=S=Q=0\right)$ (upper panel) and baryon-like  $\left(B=1,S=0,Q=1\right)$ (lower panel) Hagedorn states.}
  \label{fig:taugam}
\end{figure}

The starting point of all calculations provided here is the postulate of the statistical bootstrap model (SBM) stating that fireballs consist of fireballs which in turn consist of fireballs etc.~. As shown in \cite{Beitel:2014kza} (cf.~also \cite{Yellin:1973nj}) the mathematical formulation of this postulate 
in the formulation of \cite{Frautschi:1971ij,Hamer:1971zj}
leads to the evolving bootstrap equation for the density of states,
\begin{align}
\tau_{\vec C}(m) &= \frac{R^3}{3\pi\,m}\sum_{\vec C_1,\vec C_2}
\iint \dd m_1\dd m_2\,
m_1\tau_{{\vec C}_1}(m_1)\,m_2\tau_{{\vec C}_2}(m_2)\,\nonumber\\
&\quad\times\ p_{\rm cm}(m,m_1,m_2)\,
\delta(\vec C-\vec C_1-\vec C_2)\ ,
\end{align}
where the conservation of the quantum numbers baryon number, strangeness and electrical charge  $\vec C=(B,S,Q)$ is guaranteed. 
A major aspect was to restrict on Hagedorn states, which just consist of two constituents in order to allow for generation processes as $2\to 1$ and decay as $1\to 2$. This is some necessary constraint in order to allow for a microscopical calculation in a transport model simulation based on cross sections.
The above set of integral equations of Volterra type is solved numerically by disretizing the mass range and consecutively stepping to higher mass bins. The zeroth order input are the known hadrons according their spectral functions as implemented in the transport model UrQMD \cite{Bass:1998ca}.
The radius $R$ is the only free parameter of the model. Choosing
reasonable values as $R=0.8\fm$ ($1.0\fm$) yield Hagedorn
temperatures, i.e.~the slopes of the exponentially rising Hagedorn
spectra $\tau_{\vec C}(m)~\sim m^a\exp(m/T_H)$ as $T_H=162\MeV$
($145\MeV$) nearly independent of the quantum numbers, as indicated in fig.~\ref{fig:taugam}.

In order to treat the decay and the production of Hagedorn states, the decay width $\Gamma$ and the production cross section $\sigma=\pi R^2$ are connected via detailed balance as
\begin{align}
\Gamma_{\vec C}(m) &= \frac{\sigma}{2\pi^2\,\tau_{\vec C}(m)}\sum_{\vec C_1,\vec C_2}
\iint \dd m_1\dd m_2\,
\tau_{{\vec C}_1}(m_1)\,
\tau_{{\vec C}_2}(m_2)\,\nonumber\\
&\qquad\times\ p_{\rm cm}^2(m,m_1,m_2)\,
\delta(\vec C-\vec C_1-\vec C_2)\ ,
\end{align}
which represents the second major equation of the model.
Corresponding Hagedorn spectra and their decay width are shown in fig.~\ref{fig:taugam}.
One observes, that the slope decreases with increasing volume of the
Hagedorn state. Correspondingly also the width
decreases. Nevertheless, for larger masses, the width is nearly
constant in the range of some hundreds of MeV, and also stays finite
in the infinite mass limit.

\begin{figure}
  \centering

  \hspace*{\fill}
  \begin{minipage}{5cm}
  \includegraphics[width=5.5cm]{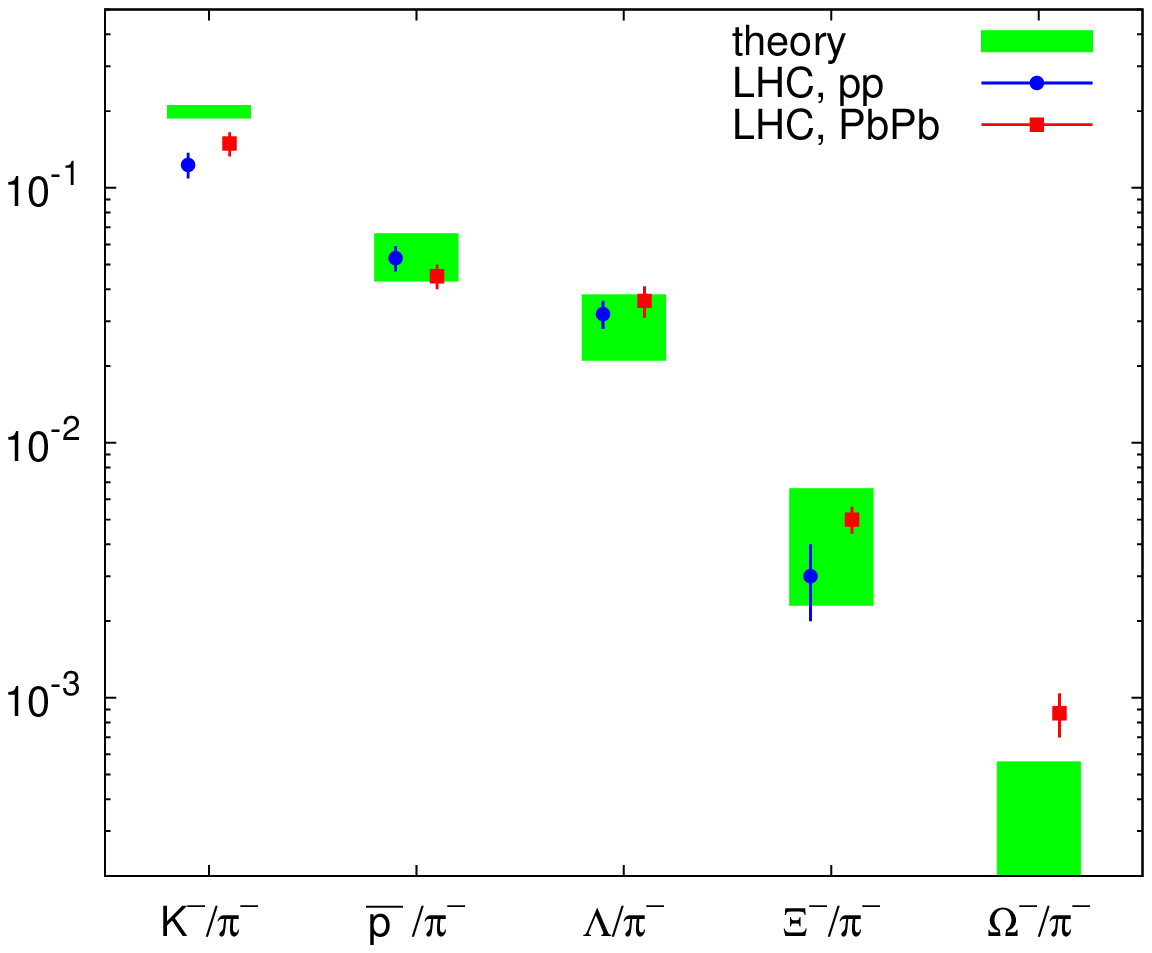}
  \end{minipage}
  \hspace*{\fill}
  \begin{minipage}{5cm}
  \includegraphics[height=4.8cm,angle=270]{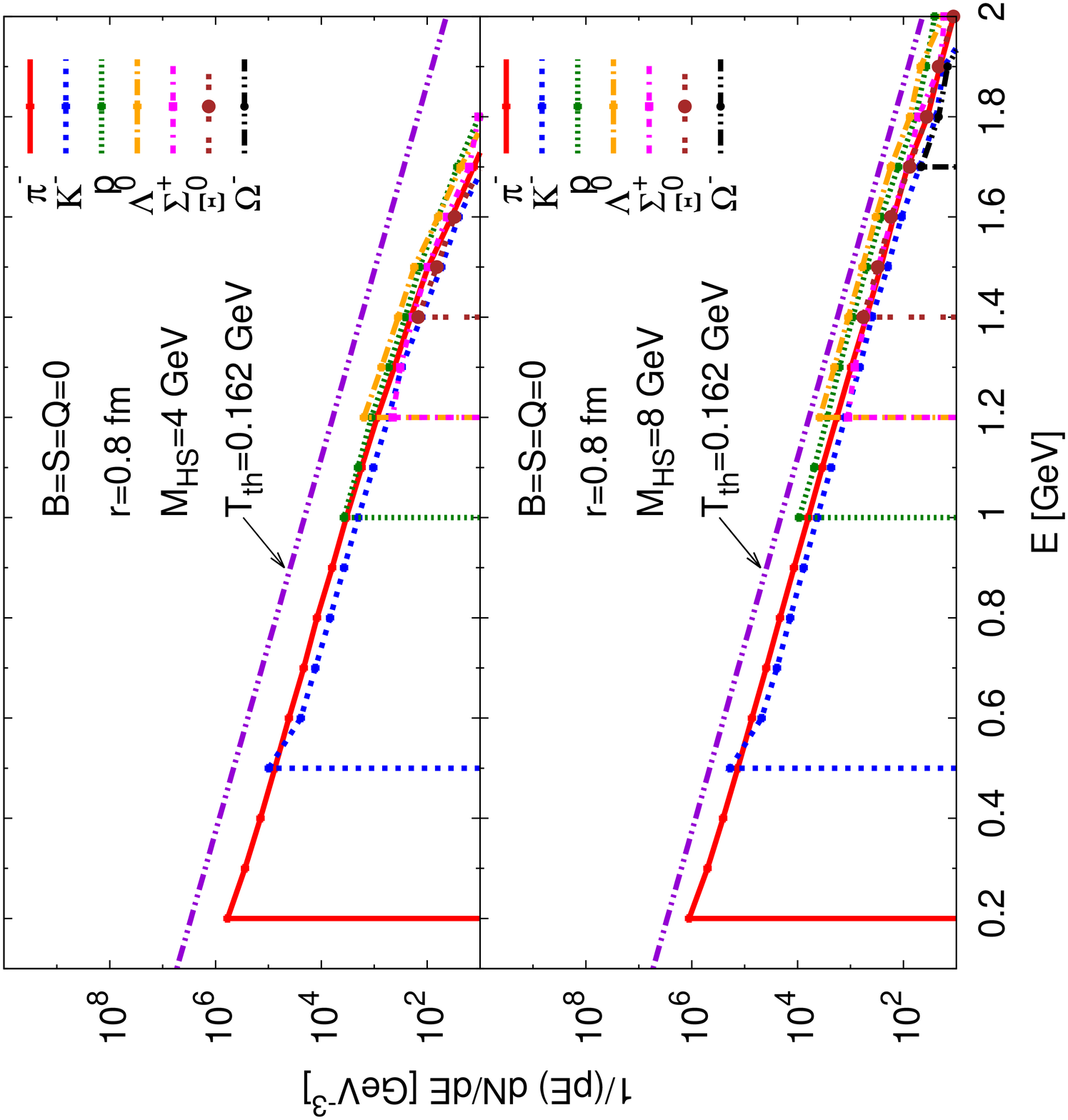}
  \end{minipage}
  \hspace*{\fill}

  \caption{Left: Comparison of particle multiplicity ratios from theory
    vs.~p-p at $\sqrt{s_{NN}}=0.9$\TeV \cite{Aamodt:2011zza} and Pb-Pb
    at $\sqrt{s_{NN}}=2.76$\TeV
    \cite{Abelev:2013vea,Abelev:2013xaa,Abelev:2013zaa}, both from
    ALICE at LHC. The bands show the range of calculations from
    cascade decay of a single charge neutral Hagedorn state with radius
    $R=0.8\fm$ and initial mass $M_{HS}=4\GeV$ and $M_{HS}=8\GeV$.
    Right: Energy spectra of hadrons stemming from cascade decay of
    charge neutral Hagedorn state with radius $R=0.8\fm$ and initial
    mass $M_{HS}=4\GeV$ and $M_{HS}=8\GeV$.}
  \label{fig:ratio}
\end{figure}

Having the total decay width and the corresponding composition, one is
able to calculate the branching ratios for the decay into light
Hagedorn states and/or hadrons. Performing such cascading decay
simulations, the resulting hadron multiplicity coincide very well with
experimentally observed values\cite{Beitel:2014kza}. This is visualized in
fig.~\ref{fig:ratio}, where the multiplicity ratios for strange and
multi-strange particles are compared with experimental values measured
at LHC.
The most striking result is the fact, that the energy spectra of
these decay hadrons \textit{look} thermal, i.e.~the functional
behavior is exponential, as shown in fig.~\ref{fig:ratio}. The slope
parameter coincides with the underlying Hagedorn temperature.

All these ingredients were implemented into UrQMD \cite{Bass:1998ca},
which can handle the creation and the decay of Hagedorn states and
their component dynamically. This enables us to perform box
calculations and study the time evolution and the equilibration of the
system. The model is able to handle different initialization
scenarios, as e.g.~the initialization with only nucleons (or pions)
with a given particle and energy density. While this may stand for some
'bottom-up' scenario, alternatively one may start with a given
Hagedorn state distribution and look into the resulting hadron
distributions. This indicates a 'top-down' scenario, where the Hagedorn
states may stem from some preceding deconfined phase.

\begin{figure}
  \centering

  \hspace*{\fill}
  \includegraphics[width=4.5cm,angle=270]{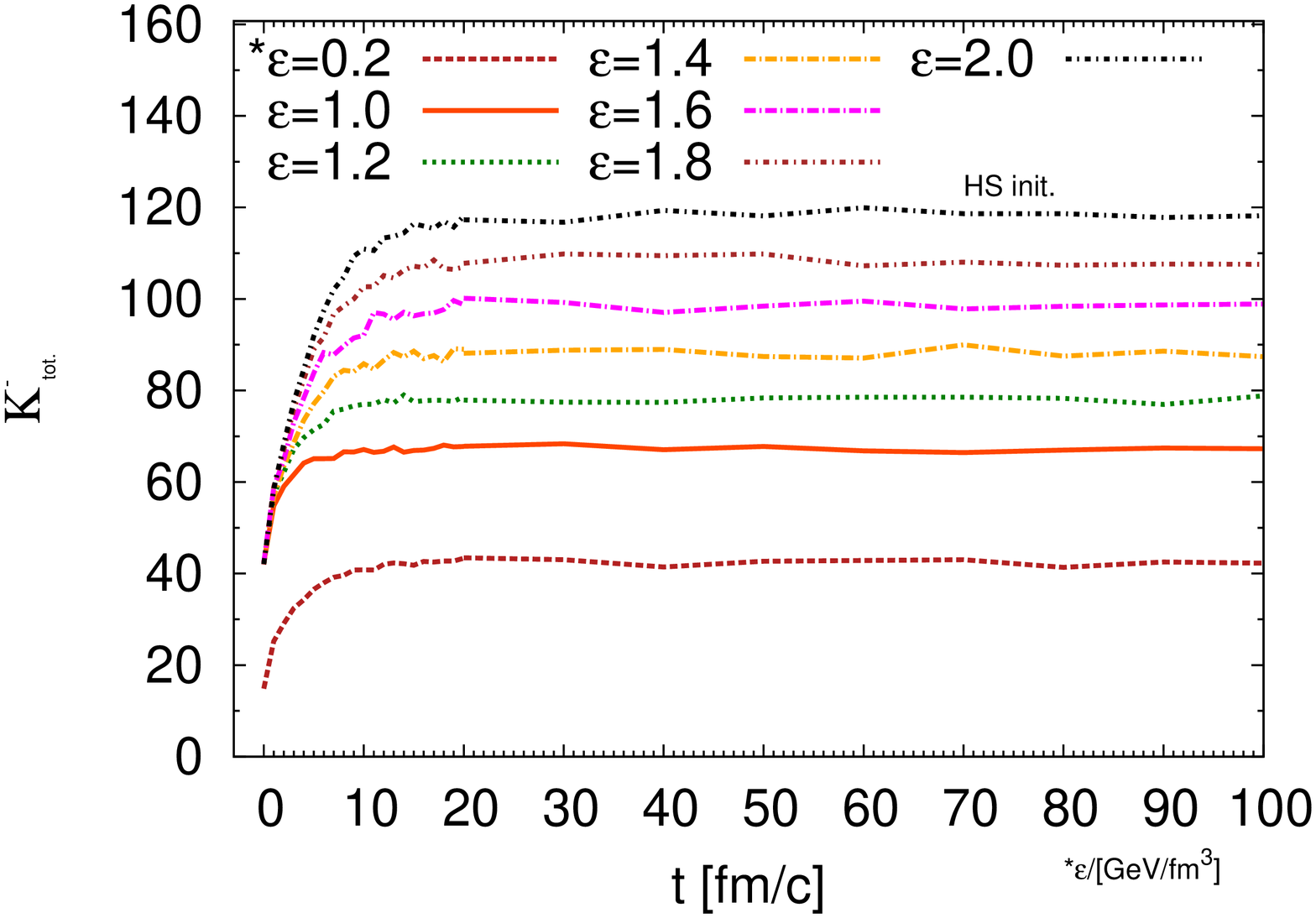}
  \hspace*{\fill}
  \includegraphics[width=4.5cm,angle=270]{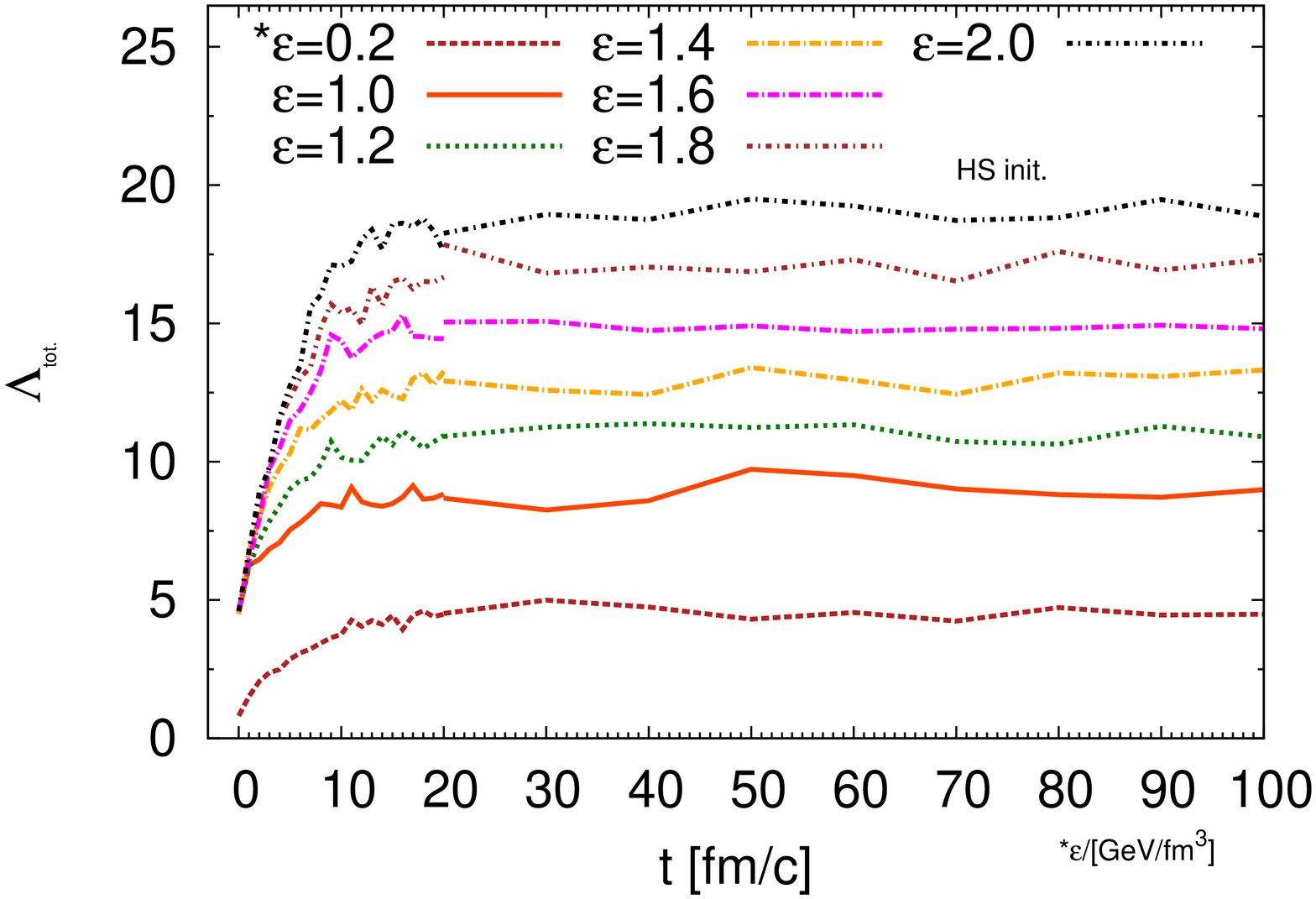}
  \hspace*{\fill}

  \caption{The time dependence of $K^-$ (left) and $\Lambda$ hadrons
    in a box calculation, initialized with Hagedorn states, for
    different initial energy densities as indicated in the plot.}
  \label{fig:multiplicities}
\end{figure}

\begin{figure}
  \centering

  \hspace*{\fill}
  \includegraphics[width=4.8cm,angle=270]{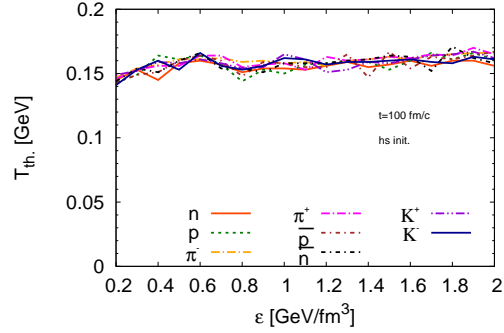}
  \hspace*{\fill}

  \caption{The slope parameter of the energy distribution of different
    particle species for different initialization energy densities.}
  \label{fig:temperature}
\end{figure}

Fig.~\ref{fig:multiplicities} shows the time
dependence of kaons and lambdas for box calculations, where only
Hagedorn states according to different energy densities were
initialized.
The given multiplcities at some time are the resulting muliplicities, if all
particles would decay down to the stable remnants.
One indeed observes very short time scales for the equilibration:
While the kaons are almost equilibrated after 2\fmc, the number of
lambdas reaches values close to the maximum value already after
5\fmc. Not shown here, but the last statement holds also for nucleons,
while pions are even faster equilibrated than the kaons. The
multiplicity ratios are almost stable for all energy densities, except
for the smallest considered one, i.e.~for $\varepsilon=0.2\GeVfmc$. 
In addition we find, that the energy spectra of the hadrons again
follow an exponential shape, with slope parameters very close to the Hagedorn temperature $T_H=162\MeV$, see fig.~\ref{fig:temperature}.


We conclude, that Hagedorn states are a valuable tool
for the understanding of the phase transition between a deconfined and a
hadronic phase. While already the decay particles of one single
Hagedorn state \textit{look} thermal with the Hagedorn temperature
as the only scale, a collective system of Hagedorn states leads to
small equilibration times of hadrons in the order of few $\fmc$. These remarkable features may also open the door for
alternative descriptions of heavy ion collisions, where directly from
a pure gluonic phase without quarks a transition to the confined phase is
possible \cite{Stoecker:2015zea}.

\vspace*{0.5cm}


\providecommand{\newblock}{}

\end{document}